\newcommand{\un}{~\mathrm}
\newcommand{\ie}{{\em i.e. }}

\documentclass[twocolumn
,floatfix
,showpacs
,preprintnumbers
,amsmath
,amssymb
,superscriptaddress]{revtex4}

\usepackage[T1]{fontenc}
\usepackage{graphicx}
\usepackage{dcolumn}
\usepackage{bm}

\begin{document}
\title{Two-dimensional scaling properties of experimental fracture surfaces}
\author{L. Ponson}
\author{D. Bonamy}
\author{E. Bouchaud}
\affiliation{Fracture Group, Service de Physique et Chimie des Surfaces et Interfaces, DSM/DRECAM/SPCSI, CEA Saclay,
F-91191 Gif sur Yvette, France} 

\begin{abstract}
The self-affine properties of post-mortem fracture surfaces in silica glass and aluminum alloy were investigated through the 2D height-height correlation function. They are observed to exhibit anisotropy. The roughness, dynamic and growth exponents are determined and shown to be the same for the two materials, irrespective of the crack velocity. These exponents are conjectured to be {\em universal}.
\end{abstract}

\date{\today}
\maketitle

Understanding the physical aspects of fracture in heterogeneous materials still presents a major challenge. Since the pioneering work of Mandelbrot \cite{Mandelbrot}, a large amount of studies have shown that crack surface roughening exhibits some universal scaling features although it results from a broad variety of material specific processes occurring at the microstructure scale (see Ref. \cite{Bouchaud4} for a review). Fracture surfaces were found to be self-affine over a wide range of length scales. In other words, the height-height correlation function $\Delta h ( \Delta r) = <[h(r+\Delta r)-h(r)]^2>_{r}^{1/2}$ computed along a given direction is found to scale as $\Delta h \sim (\Delta r)^H$ where $H$ refers to the Hurst exponent. The roughness exponent was found to be $H \approx 0.8$, weakly dependent on the nature of the material and on the failure mode \cite{Bouchaud9}. This quantity was then conjectured to be {\em universal}.

Since the early 90s, a large amount of theoretical studies suggested scenarios to explain these experimental observations. They can be classified into two main categories: (i) percolation-based models where the crack propagation is assumed to result from a damage coalescence process \cite{Roux2,Hansen}; (ii) elastic string models that consider the crack front as an elastic line propagating through randomly distributed microstructural obstacles \cite{JPBouchaud,Ramanathan}. The fracture surface corresponds then to the trace left behind this crack front.

All these models lead to self affine fracture surfaces with various exponents. However, none of them has been able to predict the measured value of the roughness exponent. The main difference between the predictions of these two categories of theoretical descriptions is that models (i) lead to isotropic fracture surfaces while models (ii), where the direction of front propagation clearly plays a specific role, predict anisotropic surfaces. The analysis of such anisotropy on experimental examples is the central point of this paper.

\begin{figure}[!h]
\includegraphics[width=0.7\columnwidth]{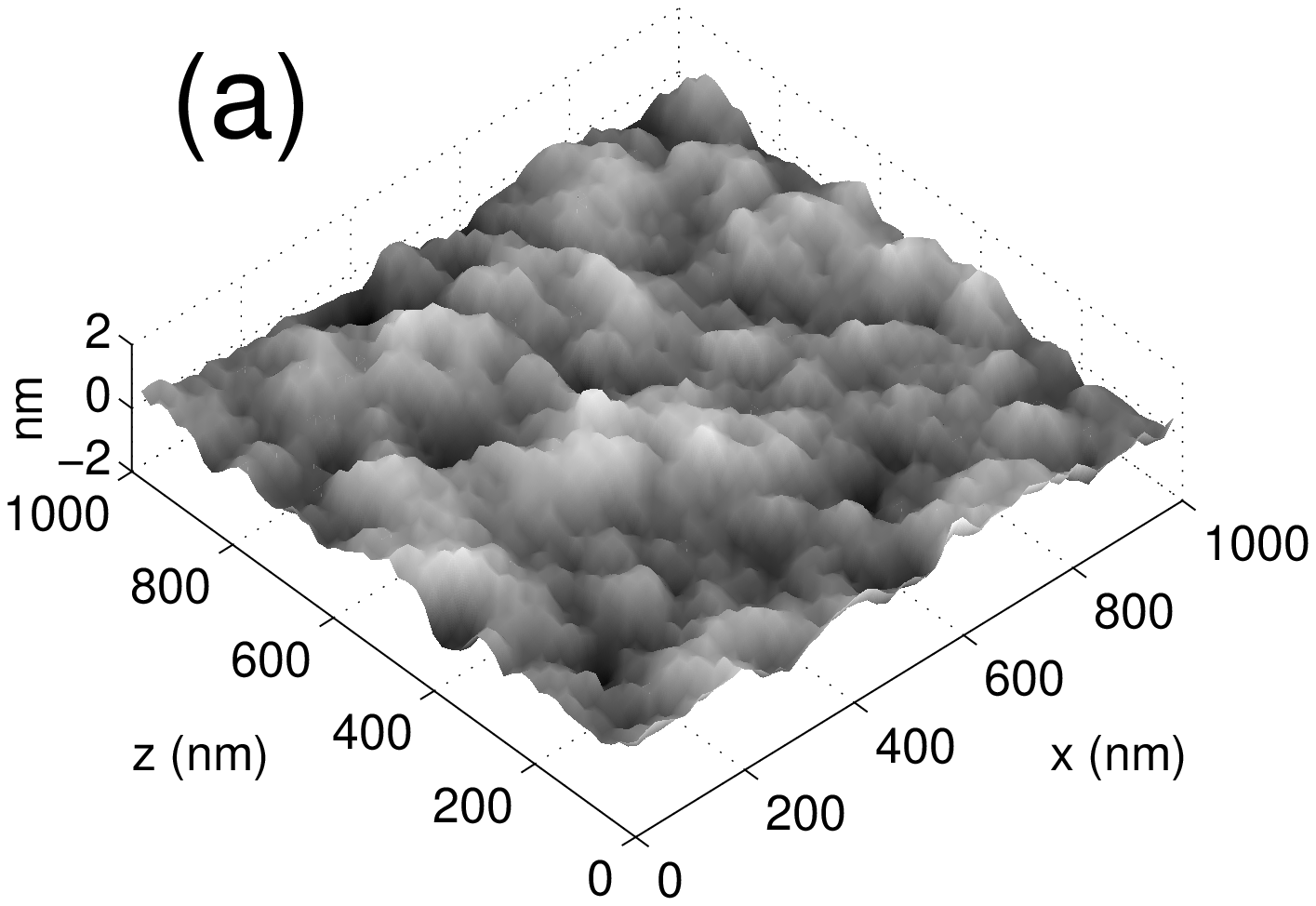}
\includegraphics[width=0.7\columnwidth]{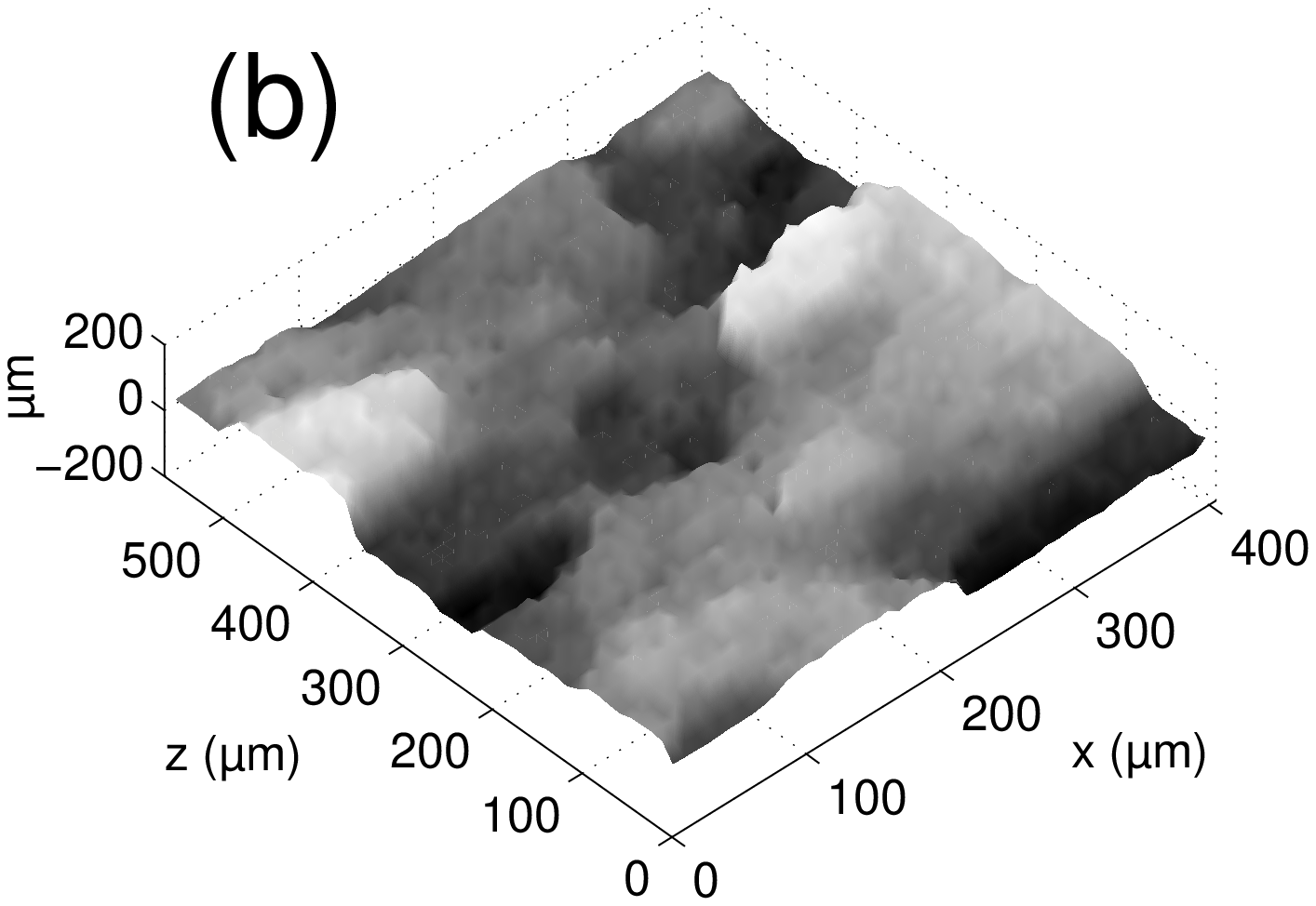}
\centering
\caption{Topographic image of fracture surface of pure silica glass (a) and aluminum alloy (b). x is the direction of crack propagation. z is parallel to the initial crack front.}
\label{fig1}
\end{figure}

We investigate the 2D scaling properties of fracture surfaces in silica glass and metallic alloy, representative of brittle and ductile materials respectively. Those were broken using various fracture tests (stress-corrosion and dynamic loading). Fracture surfaces observed for all these materials/failure modes are shown to be self-affine, in agreement with results reported in the literature. However, their scaling properties are not isotropic as usually believed but require the use of a two-dimensional (2D) height-height correlation function $\Delta h(\vec{\Delta r}) = <[h(\vec{r}+\vec{\Delta r})-h(\vec{r})]^2>_{\vec{r}}^{1/2}$ for a complete description. This 2D description involves two independent scaling exponents which correspond to the Hurst exponents measured along the crack propagation direction and the perpendicular one,  the crack front direction. They are found to vary insignificantly for the two materials and from slow to rapid crack growth. Such observations are interpreted within the framework of elastic line models driven in a random medium.

{\em Experimental setup. - }
Silica glass and a metallic alloy are chosen as the archetypes of brittle and ductile materials respectively.

Fracture of silica is performed on DCDC (double cleavage drilled compression) parallelepipedic ($5 \times 5 \times 25 \un{mm}^3$) samples under stress corrosion in mode I (see ref. \cite{Bonamy} for details). After a transient dynamic regime, the crack propagates at slow velocity through the specimen under stress corrosion. This velocity is measured by imaging in real time the crack tip propagation with an Atomic Force Microscope (AFM). In the stress corrosion regime, the crack growth velocity can be varied by adjusting properly the compressive load applied to the DCDC specimen \cite{Bonamy}. The protocol is then the following: (i) a large load is applied to reach a high velocity; (ii) the load is decreased to a value lower than the prescribed one; (iii) the load is increased again up to the value that corresponds to the prescribed velocity and maintained constant. This procedure allows us to get various crack velocities ranging from $10^{-6}$ to $10^{-11}$ m.s$^{-1}$ corresponding to zones on the post-mortem fracture surfaces which are clearly separated by visible arrest marks. The topography of these fracture surfaces is then measured through AFM with an in-plane and out-of-plane resolutions estimated of the order of $5\un{nm}$ and $0.1\un{nm}$ respectively. To ensure that there is no bias due to the scanning direction of the AFM tip, each image is scanned in two perpendicular directions and the analyses presented hereafter are performed on both images. These images represent a square field of $1\times1~\mu\mathrm{m}$ (1024 by 1024 pixels).

Fracture surfaces of the commercial 7475 aluminum alloy were obtained from CT (compact tension) specimens which were first precracked in fatigue and then broken through uniaxial mode I tension. The crack velocity varies during the fracture process, but has not been measured. In the tensile zone, the fracture surface has been observed with a scanning electron microscope at two tilt angles. A high resolution elevation map has been produced from the stereo pair using the cross-correlation based surface reconstruction technique described in \cite{Amman}. The reconstructed image of the topography represents a rectangular field of $565\times405~\mu\mathrm{m}$ (512 by 512 pixels). The in-plane and out-of-plane resolutions are of the order of $2-3~\mu\mathrm{m}$.

\begin{figure}[!h]
\includegraphics[width=0.7\columnwidth]{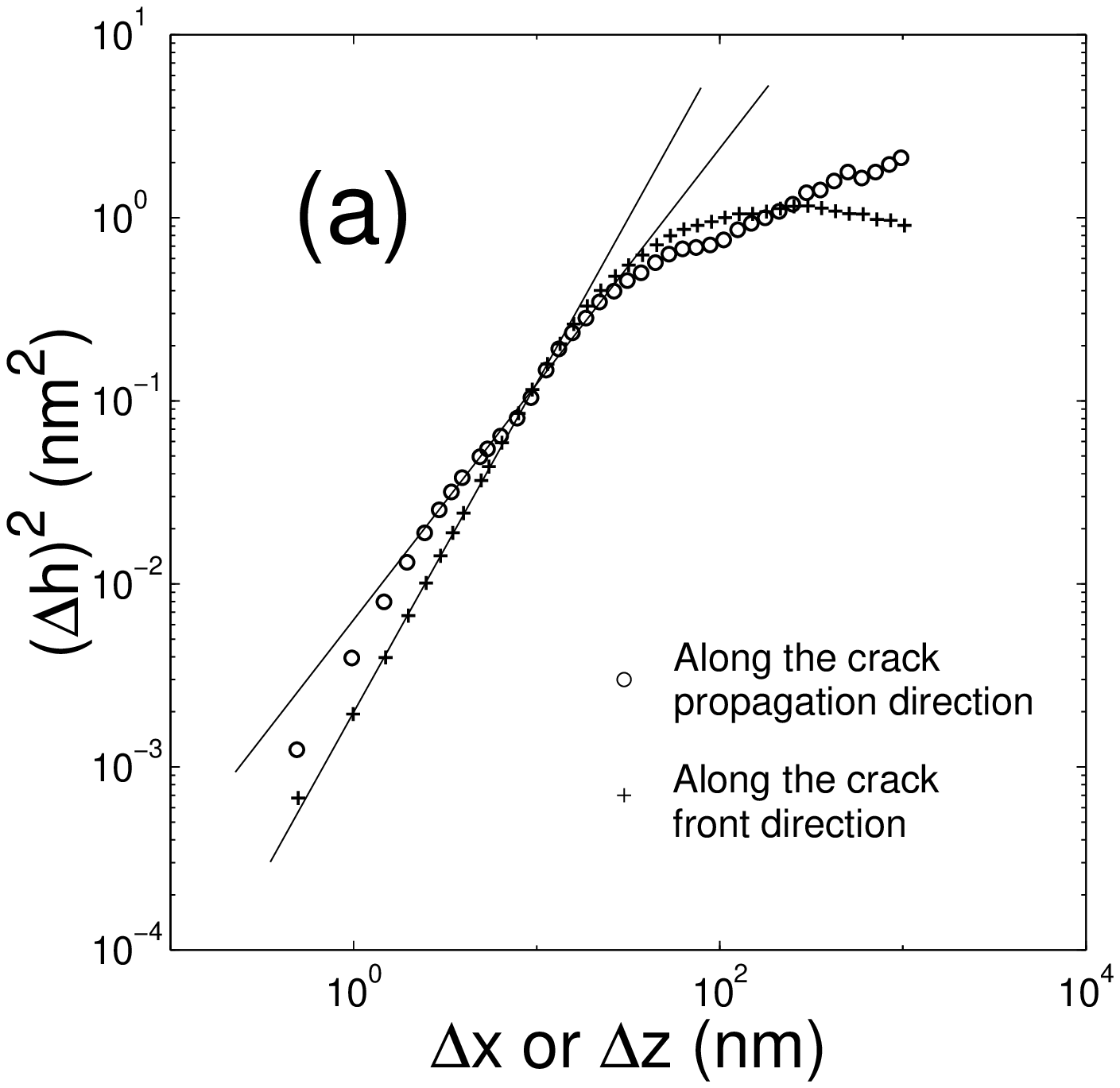}
\includegraphics[width=0.7\columnwidth]{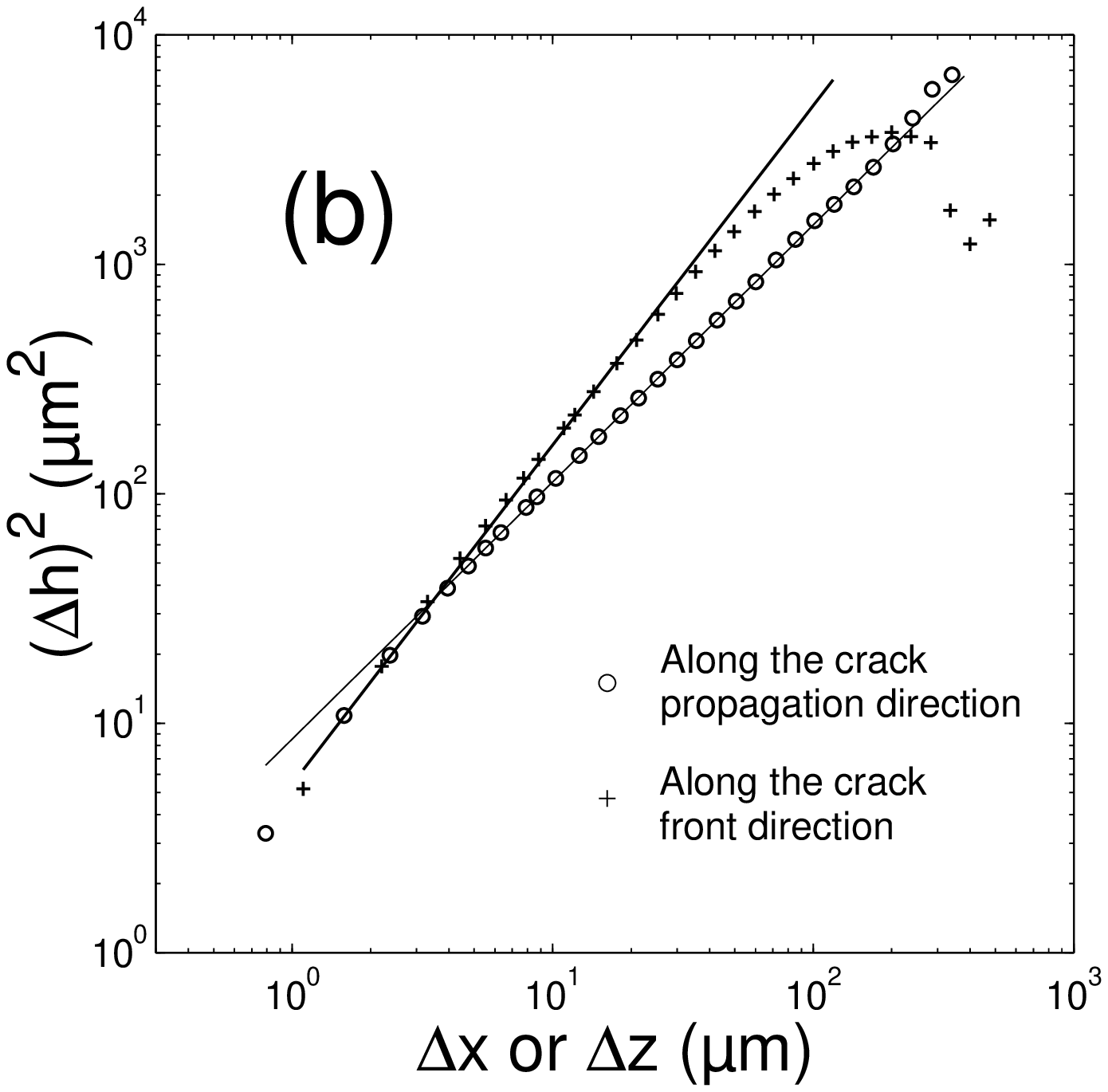}
\centering
\caption{Height-height correlation function calculated along the propagation direction and the crack front direction on a fracture surface of silica glass obtained with a crack velocity of $10^{-11}$ m.s$^{-1}$ (a) and aluminum alloy (b). The straight lines are power law fits (see text for details).}
\label{fig2}
\end{figure}

{\em Experimental results. - }
A typical snapshot of silica glass (resp. metallic alloy) fracture surface is presented in Fig. 1a (resp. Fig.1b). In both cases, the reference frame (x, y ,z) is chosen so that axis x and z are respectively parallel to the direction of crack propagation and to the crack front. The in-plane and out-of-plane characteristic length scales are respectively of the order of 50 $\mathrm{nm}$ and 1 $\mathrm{nm}$ for the silica glass, and of the order of 100 $\mu\mathrm{m}$ and 30 $\mu\mathrm{m}$ for the aluminum alloy. In order to investigate the scaling properties of these surfaces, the 1D height-height correlation functions $\Delta h(\Delta z)=<[h(z+\Delta z,x)-h(z,x)]^2>_{z,x}^{1/2}$ along the $z$ direction, and $\Delta h(\Delta x)=<[h(z,x+\Delta x)-h(z,x)]^2>_{z,x}^{1/2}$ along the $x$ direction were computed. They are represented in Fig.\ref{fig2}a (resp. Fig.\ref{fig2}b) for silica glass (resp. metallic alloy). For both materials, the profiles were found to be self-affine in both directions. Moreover, these curves indicate a clear anisotropy of the fracture surfaces. This anisotropy is reflected not only in the correlation lengths and the amplitudes but also in the Hurst exponents. Along the crack front, the exponents are found to be $0.83 \pm 0.05$ for silica and $0.75 \pm 0.03$ for metallic alloy, \ie fairly consistent with the "universal" value of the roughness exponent $\zeta \simeq 0.8$ widely reported in the literature \cite{Bouchaud9}. In the crack growth direction, the Hurst exponent is found to be significantly smaller, close to $0.63 \pm 0.04$ and $0.58 \pm 0.03$ for silica glass and the aluminum alloy respectively. In all the following, the Hurst exponent measured along the z-axis, the crack front direction, and the x-axis, the crack propagation direction, will be referred to as $\zeta$ and $\beta$ respectively. 

\begin{figure}[!ht]
\includegraphics[width=0.7\columnwidth]{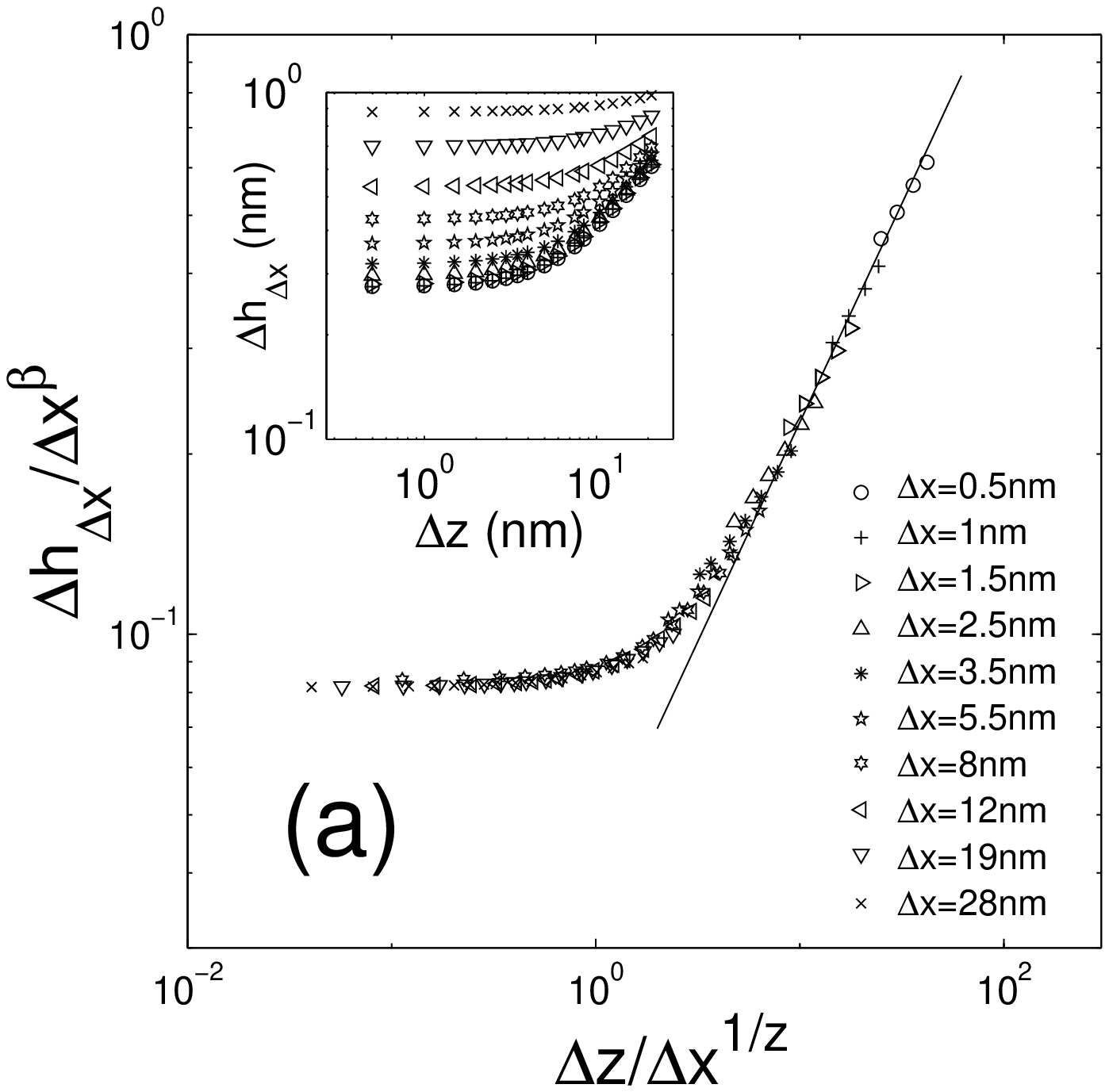}
\includegraphics[width=0.7\columnwidth]{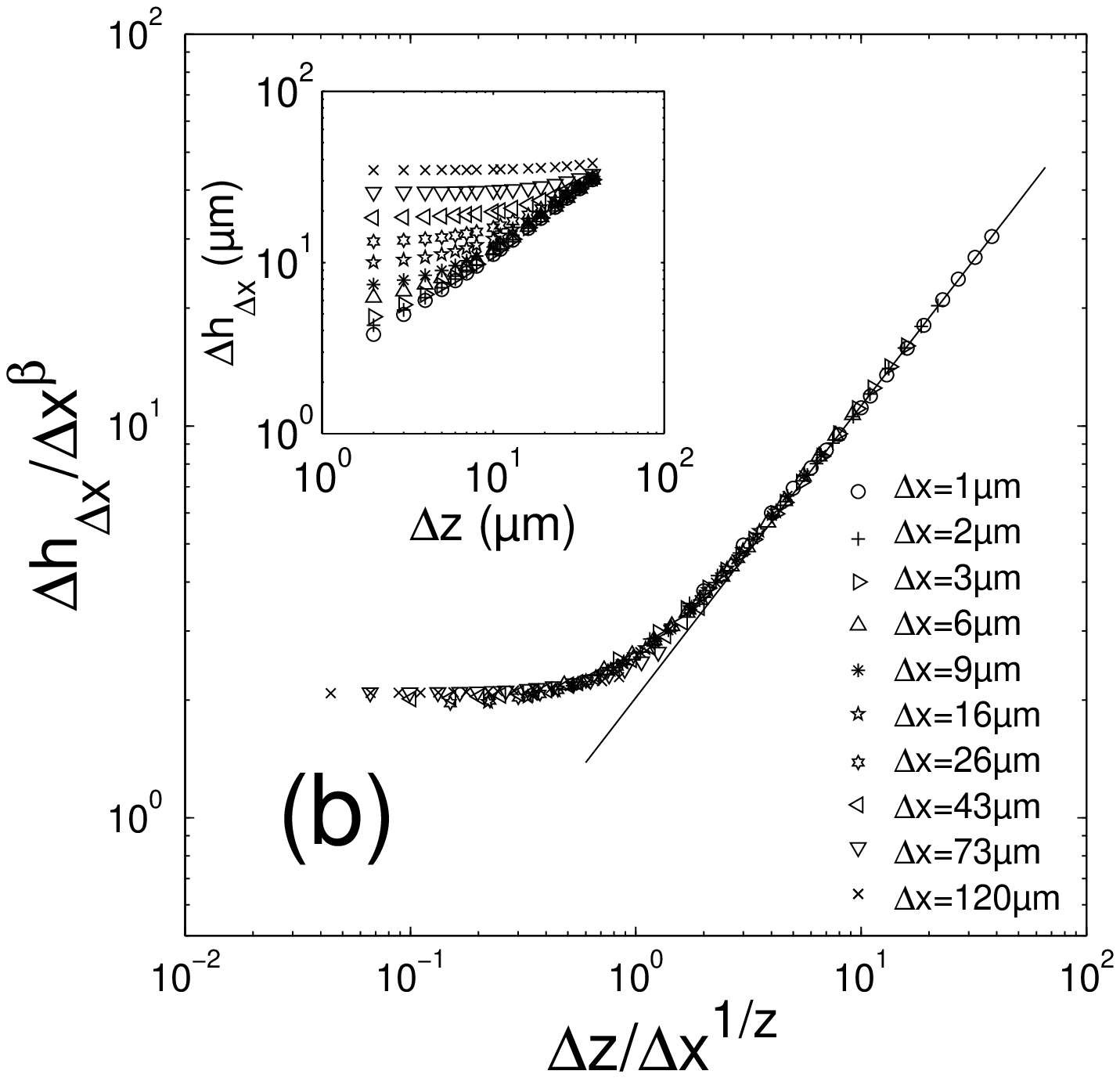}
\centering
\caption{The insets show the 2D height-height correlation functions $\Delta h_{\Delta x}(\Delta z)$ corresponding to different values 
of $\Delta x$ vs $\Delta z$ for a fracture surface of silica glass obtained with a crack velocity of $10^{-11}$ m.s$^{-1}$ (a) and aluminum alloy (b). The data collapse was obtained using Eq. \ref{cor2D} with exponents reported in Tab. \ref{tab1}.}
\label{fig3}
\end{figure}

The observation of two different scaling behaviors in two different directions of the studied fracture surfaces suggests a new approach based on the analysis of the 2D height-height correlation function defined as:

$\Delta h(\Delta z,\Delta x)=<[h(z+\Delta z,x+\Delta x)-h(x,z)]^2>^{1/2}_{z,x}$.

This function contains informations on the scaling properties of a surface in {\it all} directions. The variations of the correlation functions $\Delta h_{\Delta x}$ are plotted as a function of $\Delta z$ in the insets of Figure \ref{fig3}a and \ref{fig3}b for silica glass and aluminum alloy fracture surfaces respectively. For adequate values of $\beta$ and $z$, it can be seen in the main graphs of that same figure that a very good collapse of the curves can be obtained by normalizing the abscissa and the ordinate by $\Delta x^{1/z}$ and $\Delta x^{\beta}$ respectively. The resulting master curve is characterized by a plateau region and followed by a power law variation with exponent $\zeta$. In other words:

\begin{equation}
\begin{array} {l}
   \Delta h(\Delta z,\Delta x)=\Delta x^{\beta}f(\Delta z/\Delta x^{1/z}) \\
\\
$where$\quad	f(u) \sim \left\{
\begin{array}{l l}
1 & $if u$ \ll 1  \\
u^{\zeta} & $if u$ \gg 1
\end{array}
\right.
\end{array}
\label{cor2D}
\end{equation}

The exponents $\beta$ and $z$ which optimize the collapse, and the $\zeta$ exponent determined by fitting the large scales regime exhibited by the master curves are listed in Table \ref{tab1}. The three exponents are found to be $\zeta \simeq 0.75$, $\beta \simeq 0.6$ and $z \simeq 1.25$, independent of the material and of the crack growth velocity over the whole range from ultra-slow stress corrosion fracture (picometer per second) to rapid failure (some meters per second). The ratio of $\zeta$ to $\beta$ is given in the fourth column of Table \ref{tab1}. It is worth to note that the exponent z fulfills the relation $z=\zeta/\beta$. The same exponents have also been observed on fracture surfaces of mortar and wood \cite{Ponson4}. They are therefore conjectured to be {\em universal}.

\begin{table}[!h]
\begin{center}
\begin{tabular}{|l|c|c|c|c|c|}
 \hline
  & $\zeta$ & $\beta$ & z & $\zeta/\beta$
   \\ \hline \hline silica glass & 0.77 $\pm$ 0.03 & 0.61 $\pm$ 0.04 & 1.30 $\pm$ 0.15 & 1.26
   \\ \hline metallic alloy & 0.75 $\pm$ 0.03 & 0.58 $\pm$ 0.03 & 1.26 $\pm$ 0.07 & 1.29
   \\ \hline
\end{tabular}
\caption{Scaling exponents measured on fracture surfaces of silica glass and metallic alloy. $\zeta$, $\beta$, $z$ and $\zeta/\beta$ are respectively interpreted as the roughness exponent, the growth exponent and the dynamic exponent $z$ while the fourth column contains the ratio of $\zeta$ to $\beta$. Error bars represent an interval of confidence of 95 \%.}
\label{tab1}
\end{center}
\end{table}

{\em Discussion. - }
The experiments reported in this letter explored the 2D scaling properties of fracture surfaces of two different materials. Three main conclusions can be drawn: (i) 1D profiles scanned parallel to the crack front direction and to the direction of crack propagation both exhibit self affine scaling properties, but those are characterized by two different Hurst exponents referred to as $\zeta$ and $\beta$ respectively; (ii) the 2D height-height correlation function is shown to collapse on a single curve (Eq. \ref{cor2D}) when appropriatly rescaled. This scaling involves three exponents $\zeta$, $\beta$ and $z$; (iii) These three exponents are independent of both the material considered and the crack growth velocity over the explored range.

These conclusions enables a discussion on the various competing models developed to capture the scaling properties of fracture surfaces \cite{Roux2,Hansen,JPBouchaud,Ramanathan}. The anisotropy clearly evidenced in the scaling properties of fracture surfaces cannot be captured by static models like percolation-based models \cite{Hansen}. On the other hand, these results are reminiscent to what is observed in kinetic roughening models \cite{Barabasi}. These models consider the time evolution of an elastic manifold driven in a random medium. The roughness development of the line $h(z,t)$ starting from an initially straight line $h(z,t=0)=0$ is then  characterized by a 1D height-height correlation function $\Delta h(\Delta z,t)$ that scales as \cite{Barabasi}:

\begin{equation}
\begin{array} {l}
    \Delta h(\Delta z,t)=t^ {\beta} g(\Delta z/t^{1/z}) \\
\\
$where$\quad	g(u) \sim \left\{
\begin{array}{l l}
u^{\zeta} & $if u$ \ll 1  \\
1 & $if u$ \gg 1
\end{array}
\right.
\end{array}
\end{equation}

\noindent where $\zeta$, $\beta$ and $z$ refer to the roughness, growth and dynamic exponents respectively. Signature of this roughnening scaling can also be found in the steady state regime reached at long times when the roughness becomes time invariant. In this regime, the 2D height-height correlation function $\Delta h(\Delta z,t)$ is expected to scale as \cite{Kardar2}:

\begin{equation}
\begin{array} {l}
   \Delta h(\Delta z,\Delta t)=\Delta t^{\beta}f(\Delta z/\Delta t^{1/z}) \\
\\
$where$\quad	f(u) \sim \left\{
\begin{array}{l l}
1 & $if u$ \ll 1  \\
u^{\zeta} & $if u$ \gg 1
\end{array}
\right.
\end{array}
\end{equation}

\noindent which is exactly the scaling (\ref{cor2D}) followed by the experimental surfaces after time $t$ has been replaced by coordinate $x$ measured along the crack propagation direction. This provides a rather strong argument in favour of models like \cite{JPBouchaud,Ramanathan} that describe the fracture surface as the juxtaposition of the successive crack front positions - modelled as a pseudo elastic line - moving through materials with randomly distributed local toughness. In this scenario, the hurst exponents $\zeta \simeq 0.75$ and $\beta \simeq 0.6$ measured along the crack front direction and the direction of crack propagation respectively coincide with the roughness exponent and the growth exponent as defined within the framework of elastic string models \cite{Barabasi}. Let us note moreover that in such models, the dynamic exponent $z$ is expected to be related to $\zeta$ and $\beta$ through $z=\zeta/\beta$ \cite{Family}. This leads to a value of $z=1.25$ in perfect agreement with the value measured experimentally. 

In elastic line models, the set of exponents $\zeta$, $\beta$ and $z$ depends only on the dimensionality \cite{Barabasi, Chauve}, the range of the elastic interaction \cite{Tanguy, Chauve} and, to some extent, on the line velocity \cite{Leschhorn2}. It has been shown that for crack propagating in a linear elastic solid, the restoring elastic forces are long range rather than local \cite{Gao}. Corresponding elastic line models predict logarithmic correlations \cite{Ramanathan}, which is significantly different from $\zeta \simeq 0.75$ as reported in this paper. Let us note that the same model applied to the interfacial crack problem leads to a roughness exponent $\zeta \simeq 0.39$ \cite{Rosso2} and $z \simeq 0.75$ \cite{Tanguy, Schmittbuhl4}, while experiments \cite{Schmittbuhl3} reported values $\zeta \simeq 0.6$ and $z \simeq 1.2$. These experimental values are much closer to the ones expected in elastic line models with {\em short} range elastic interactions, that predict roughness exponents $\zeta \simeq 0.63$ \cite{Rosso}. Understanding the origin of the interaction screening in crack problems provides a significant challenge for future investigation.

Finally, it is worth to mention that the scaling properties exhibited by fracture surfaces may have interesting expertise applications. It allows indeed to determine the direction of crack propagation from the analysis of post-mortem fracture surfaces and, thus to reconstruct the history of the processes that have lead to the failure of the structure.

\begin{acknowledgments}
We thank H. Auradou, J. P. Bouchaud, E. Bouchbinder, C. Guillot, J. P. Hulin, G. Mourot, S. Morel, I. Procaccia and S. Sela for enlightening discussions.
\end{acknowledgments}


\end{document}